\documentclass[prb,twocolumn,showpacs,floats,eqsecnum,amsmath,amssymb,nofootinbib]{revtex4}
\usepackage[dvips]{graphicx}

\begin{document}



\title{Charge density plateaux and insulating phases in the $t-J$ model 
with ladder geometry}

\author{A. Fledderjohann$^{1}$, A. Langari$^{1,2}$ and K.-H. M\"utter$^{1}$}

\affiliation{$^1$Physics Department, University of Wuppertal, 42097 Wuppertal,
Germany}
\affiliation{
$^2$Institute for Advanced Studies in Basic Sciences,
Zanjan 45195-159, Iran}


\begin{abstract}
\leftskip 2cm
\rightskip 2cm
We discuss the occurrence and the stability of charge density plateaux in
ladder-like $t-J$ systems (at zero magnetization $M=0$) for the cases 
of 2- and 3-leg ladders. Starting from isolated rungs at zero leg coupling,
we study the behaviour of plateaux-related phase transitions by means of
first order perturbation theory and compare our results with Lanczos 
diagonalizations for $t-J$ ladders ($N=2\times 8$) with increasing leg couplings.
Furthermore we discuss the regimes of rung and leg couplings that should be
favoured for the appearance of the charge density plateaux.


\end{abstract}


\maketitle



\section{Introduction}

The $t-J$ model has been introduced as the first order correction
of the extreme atomic limit of the Hubbard model \cite{harris67,
brinkman70} and is considered the simplest model including the low
energy physics of doped ladder systems. \cite{dagotto96,rice97}
The phase diagram of the two leg Hubbard model has been
investigated in [\onlinecite{balents96}] by means of a 
renormalization group approach valid for small values of the
on-site Coulomb interaction $U$ but for arbitrary charge density
and arbitrary hopping along the rungs. 
The phase diagram is classified as different $C_xS_y$ phases which denotes
$x$ gapless charge modes and $y$ gapless spin modes. They have shown under which
condition a phase of $C_1S_0$ appears which is analog of either a superconductor
or charge-density wave.
The extension of this
approach to the $N$-leg Hubbard model can be found in
[\onlinecite{lin97}]
where the dimensional crossover as $N\rightarrow\infty$ is discussed.
The charge and spin gap for the two leg Hubbard model and its
dependence on the on-site Coulomb interaction and the rung
hopping parameter $t$ have been calculated by means of the
density matrix renormalization group (DMRG)
and compared with the previous weak-coupling RG.\cite{park99}
The effect of an additional nearest neighbour Coulomb repulsion
$V$ has been studied in [\onlinecite{vojta01}].
A charge order (metal-insulator) transition was found at charge 
density $\rho=1/2$ from a homogeneous state to a charge density
wave. The influence of an anisotropy between leg and rung
couplings in Hubbard and $t-J$ models on specific correlations,
which signal the metal-insulator transition, has been investigated
in [\onlinecite{riera99}].
The metal-insulator transition is accompanied by the opening
of a gap, which appears as a plateau in the charge density
$\rho(\mu)$ as a function of the chemical potential $\mu$.
The charge density plateau in $\rho(\mu)$ looks similar to the
plateaux in the magnetization curve $M(B)$ found in the spin
ladder systems and one might ask whether the mathematical
foundations for both plateaux are the same. This is indeed
the case and becomes evident if one maps the $t-J$ Hamiltonian
on a spin-1 Hamiltonian with broken $SU(3)$ symmetry.\cite{fle03}
The Lieb-Schultz-Mattis theorem\cite{lieb61} has been extended to quasi
onedimensional fermionic systems \cite{giagliardini98} and the
momenta of low-lying excitations could be classified thereby.
The quantization rule of Oshikawa, Yamanaka and Affleck
\cite{oshikawa97} predicts
as well possible plateaux in the charge density $\rho(\mu)$.

The values of the charge density (and magnetization) at the
plateaux are fixed by the geometry of the system (e.g. the
number of legs in a ladder system).

\begin{figure}[ht!]
\centerline{\includegraphics[width=8.0cm,angle=0]{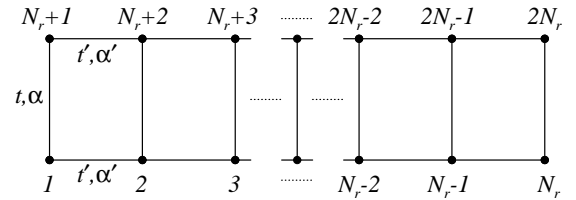}}
\caption{Structure of a two leg ladder with open boundary conditions
and $2N_r$ sites.}
\label{fig1}
\end{figure}

In this paper we will study the ground states of the $t-J$
Hamiltonian on a ladder with $n$ legs ($n=2,3$)
\begin{eqnarray}
H^{[n]} & = & t H_r^{[n]} + t' H_l^{[n]}\,.\label{eq1.1}
\end{eqnarray}
$t,t'$ are the hopping parameters, and
\begin{eqnarray}
H_r^{[n]} & = & \sum_{x=1}^{N_r} h_r^{[n]}(x,\alpha),
\quad \alpha=J/t\,,\nonumber\\
 & & \label{eq1.2}\\
H_l^{[n]} & = & \sum_{x=1}^{N_r-1}h_l^{[n]}(x,\alpha')\,
\quad \alpha'=J'/t'\nonumber
\end{eqnarray}
define the contributions of the couplings along the rungs and 
legs, respectively. The spin exchange is included in the ratios
$\alpha=J/t$, $\alpha'=J'/t'$.
The latter are depicted in Figs. \ref{fig1} and \ref{fig2} for
the cases of a two and a three leg ladder.

A charge density plateau in $\rho(\mu)$ -- where $\mu$ is the
chemical potential -- is signalled by discontinuous changes
in the slope of the ground state energy per site as function
of $\rho$ [cf. eq. (\ref{eq2.14})].
They emerge immediately in the ``local rung approximation''
\cite{riera99} (or ``bond operator theory''\cite{park01}),
where the ground states are direct products of rung cluster
states. The corresponding energies turn out to be piecewise
linear in $\rho$.

It is the purpose of this paper to go beyond the local
rung approximation by means of a systematic perturbation 
theory in the leg coupling $t'$. To first order, this leads
to an effective interaction between the rung cluster states.
The corresponding effective Hamiltonians are defined on a
chain with $N_r$ sites -- where $N_r$ is the number of rungs --
and can be diagonalized numerically.

The outline of the paper  is as follows:
In Section \ref{sec2} we first treat the two leg ladder.
Here, the effective Hamiltonians for $\rho<1/2$ and $\rho>1/2$
look like modified $t-J$ models on a chain with $N_r$ sites.

In Section \ref{sec3} we extend all considerations to the
three leg ladder,where plateaux appear at $\rho=1/3$ and
$\rho=2/3$. First order perturbation theory in $t'$ leads
to three different effective Hamiltonians in the regimes
$0\leq\rho\leq 1/3$, $1/3\leq\rho\leq 2/3$, $2/3\leq\rho\leq 1$.

In Section \ref{sec4} we compare the perturbative results with
Lanczos diagonalizations on ladder systems and discuss
consequences for the phase diagram.

\section{t-J model on a two leg ladder\label{sec2}}

According to the notation given in Fig. \ref{fig1} the couplings
along the rungs and legs which enter into the $t-J$-Hamiltonian
on a two leg ladder are given by:
\begin{eqnarray}
h_r^{[2]}(x,\alpha) & = & h(x,x+N_r,\alpha)\,,\label{eq2.1}\\
h_l^{[2]}(x,\alpha') & = & h(x,x+1,\alpha')\label{eq2.2}\\
 & & + h(x+N_r,x+1+N_r,\alpha')\,.
\nonumber
\end{eqnarray}

For our purposes it is convenient to represent these couplings
in terms of ``constrained'' permutation operators
\begin{equation}
h(x,y,\tilde{\alpha})=-P_{01}(x,y)+\frac{\tilde{\alpha}}{2}Q_{11}(x,y)
\label{eq2.4}
\end{equation}
Here, $P_{01}(x,y)$ is a permutation operator, which permutes the states at
$x$ and $y$, if they are occupied by one hole (0) and one electron (1).
This operator allows the hopping of electrons and holes and forbids the
double occupancy of each site $x$ with two electrons.

If both sites $x$ and $y$ are occupied with electrons, the operator
\begin{eqnarray}
Q_{11}(x,y) & = & P(x,y)-1,\\
Q_{11}(x,y) |x,y\rangle & = & |y,x\rangle-|x,y\rangle
\label{eq2.5}
\end{eqnarray}
first interchanges the electrons at $x$ and $y$; afterwards the original
state is subtracted. 

\subsection{0th order perturbation theory in $t'$}

The lowest energy eigenstates of the rungs $(x,x+N_r)$ with Hamiltonian
$h(x,x+N_r,\alpha)$ can be easily calculated.
\begin{eqnarray}
|Q(x)=0\rangle & = & |0,0\rangle\nonumber\\
|Q(x)=1\rangle & = & \frac{1}{\sqrt{2}}
(|\sigma,0\rangle+|0,\sigma\rangle)
\label{eq2.8}\\
|Q(x)=2\rangle & = & \frac{1}{\sqrt{2}}(|+,-\rangle-|-,+\rangle)\nonumber
\end{eqnarray}
The right-hand side defines the states on the sites $x$ and $x+N_r$
of the rung; $0$ means {\it hole} with charge zero. $Q(x)=1$ means
one electron with spin $\sigma=\pm1$. $Q(x)=2$ represents an electron
pair coupled antiferromagnetically to a total spin 0. 
Therefore, we have charges $0, 1$ and $2$
for $|Q(x)=0\rangle$, $|Q(x)=1)\rangle$ and 
$|Q(x)=2\rangle$ respectively.
The eigenvalues are given by the following equations.
\begin{equation}
h(x,x+N_r,\alpha) |Q(x)\rangle=\varepsilon_n |Q(x)\rangle
\label{eq2.10}
\end{equation}
\begin{equation}
\varepsilon_0=0; \varepsilon_1=-1; \varepsilon_2=-\alpha
\label{eq2.11}
\end{equation}
In the limit of vanishing leg couplings $t'=0$, the system of decoupled rungs
has the following ground state.
\vspace{5mm}

(1) For $\rho=\frac{Q}{2N_r}=\frac{1}{2}-\frac{q}{N_r}\leq\frac{1}{2}$;
$q=0,1,2,\dots$
and $\alpha <2$ the ground state is a direct product of $(N_r-2q)$ rung
states $|Q(x)=1(\sigma)\rangle$ and $2q$ states with $|Q(x)=0\rangle$.

Note that for $\alpha <2$, the creation of a pair state $|Q(x)=2\rangle$
and a charge zero state $|Q(x)=0\rangle$ from two charge $1$ states
$|Q(x)=1(\sigma=+1)\rangle$ and  $|Q(x)=1(\sigma=-1)\rangle$ is not
energetically favorable, since
\begin{equation}
\varepsilon_0 + \varepsilon_2 > 2 \varepsilon_1
\label{eq2.11b}
\end{equation}
Therefore, the ground state energy turns out to be
\begin{eqnarray}
E_1^{(0)}\left(\rho,\alpha\right) & = &
(N_r-2q)\varepsilon_1+2q\varepsilon_0\nonumber\\
 & = & -2N_r\rho
\label{eq2.12}
\end{eqnarray}

(2) For $\rho=\frac{Q}{2N_r}=\frac{1}{2}+\frac{q}{N_r}\geq\frac{1}{2}$;
$q=0,1,2,\dots$
and $\alpha <2$ the ground state is a direct product of $(N_r-2q)$ rung
state $|Q(x)=1\rangle$ and $2q$ states $|Q(x)=2\rangle$ with electron pairs.
In this case the ground state energy is
\begin{eqnarray}
E_2^{(0)}\left(\rho,\alpha\right) & = & 
(N_r-2q)\varepsilon_1+2q\varepsilon_2\nonumber\\
 & = & 2N_r\rho(1-\alpha)-(2-\alpha)N_r\nonumber\\
 & & \label{eq2.13} 
\end{eqnarray}
Note, in both regimes $\rho\leq 1/2$ and $\rho\geq 1/2$
the ground state energy is linear in $\rho$.
The first derivative of the ground state energy
\begin{equation}
\mu=\frac{d}{d\rho}\frac{E}{2N_r}
\label{eq2.14}
\end{equation}
is related to the chemical potential which has a discontinuity
at $\rho=\frac{1}{2}$.
\begin{eqnarray}
\mu=
\biggl\{
    \begin{array}{c}
       -1 \hspace{5mm} \mbox{for} \hspace{5mm} \rho\leq\frac{1}{2}\\
       1-\alpha \hspace{5mm} \mbox{for} \hspace{5mm} \rho\geq\frac{1}{2} ,\\
    \end{array}
\label{eq2.15}
\end{eqnarray}
This is the first indication of a charge density plateau at $\rho=1/2$
in the $t-J$ model on a two leg ladder.

To zeroth order in the leg coupling ($t'$), the eigenstates of the t-J
Hamiltonian (\ref{eq1.1}) are product states of $N_r$ rung states:
\begin{equation}
|Q(1),Q(2),\dots,Q(N_r)\rangle= 
\prod_{x}|Q(x)\rangle
\label{eq3.1}
\end{equation}
The rung quantum numbers $Q(x)=0,1(\sigma=\pm1),2$ are subjected to the conservation
of total charge $Q$,
\begin{equation}
\sum_{x} Q(x)=Q
\label{eq3.2}
\end{equation}
Moreover, the total spin, which originates from the spin of charge $1$ states
$|Q(x)=1\rangle$, is assumed to be zero here:
\begin{equation}
\sum_{x} \sigma(x) \delta_{Q(x),1}=0
\label{eq3.3}
\end{equation}
Therefore the ground states with energies (\ref{eq2.12}), (\ref{eq2.13})
are highly degenerate for $\rho\neq 0,1/2,1$.

\subsection{1st order perturbation theory in $t'$
}
A first order perturbation theory demands a computation
of the transition matrix elements:
\begin{eqnarray}
\langle Q'(1),Q'(2),\dots,Q'(N)|H_l^{[2]}|Q(1),Q(2),\dots,Q(N)\rangle = 
\nonumber \\
\sum_{x=1}^{N_r-1} {\cal A}\langle Q'(x),Q'(x+1)|h_l^{[2]}(x,\alpha')
|Q(x),Q(x+1)\rangle
\nonumber\\
 & & \label{eq3.4}
\end{eqnarray}
where $${\cal A}=\prod_{y\neq x,x+1}\delta_{Q'(y),Q(y)},$$
and the diagonalization of the resulting effective Hamiltonian on
a chain with $N_r$-sites. The matrix elements of the leg operators
between the rung states Eq.(\ref{eq2.8}) 
\begin{eqnarray}
\langle Q'(1),Q'(2)|h_l^{[2]}(1,\alpha')|Q(1),Q(2)\rangle & 
\equiv & \nonumber\\[2pt]
\left(Q'(1),Q'(2);Q(1),Q(2)\right)\,,\quad\quad & &
\label{eq3.4b}
\end{eqnarray}
-- the
explicit spin dependence of the electrons has been omitted here --
are listed in the following equations:
\begin{eqnarray}
(0,0;0,0) & = & 0\,, \hspace{0.75cm};\hspace{0.4cm} 
(\sigma,0;\sigma,0)=0\,,\label{eq3.5}\\
\hspace{-0.4cm}(\sigma'_1,\sigma'_2;\sigma_1,\sigma_2) & = &
\frac{\alpha'}{4}\langle\sigma'_1,\sigma'_2|P(1,2)-1|\sigma_1,\sigma_2\rangle
\,,\label{eq3.7}\\
(\sigma,0;0,\sigma) & = & -1\,, \hspace{0.5cm};\hspace{0.4cm}
(\sigma,2;2,\sigma)=-\frac{1}{2}\,,
\label{eq3.8}\\
(2,\sigma;2,\sigma) & = & -\frac{\alpha'}{4}\,, \hspace{0.33cm};\hspace{0.4cm}
(2,2;2,2)=-\frac{\alpha'}{2}\,.
\label{eq3.10}
\end{eqnarray}
From these equations we can read off the effective Hamiltonian on a chain
with $N_r$ sites. Since the ground state structure differs for
$\rho\leq\frac{1}{2}$ and $\rho\geq\frac{1}{2}$
as described in Sec.(II), we have to consider these two cases separately.
\vspace{5mm}

(1) For $\rho\leq 1/2$ only rung states with $Q(x)=0$
and $Q(x)=1$ are involved. From Eqs.
(\ref{eq3.5}) and (\ref{eq3.7})  we see that the effective Hamiltonian
\begin{eqnarray}
\hat{H}_1(\alpha') & = &  \sum_{x=1}^{N_r-1}\hat h_1(x,x+1,\alpha')
\label{hat_h1}
\end{eqnarray}
with couplings:
\begin{equation}
\hat h_1(x,x+1,\alpha')=-P_{01}(x,x+1)+\frac{\alpha'}{4}Q_{11}(x,x+1)
\label{eq3.11}
\end{equation}
can be identified with a $t-J$ model (\ref{eq2.4}) on a chain with coupling
$\alpha'/2$, which is just half of the leg coupling.
Note also that the charge density on the chain with $N_r$ sites
\begin{equation}
\rho_1=\frac{Q}{N_r}=2\rho
\label{eq3.12}
\end{equation}
is just twice the charge density of the two leg ladder.
In first order perturbation theory we get for the ground state energy
on the two leg ladder:
\begin{eqnarray}
E\left(\rho, t=1, \alpha; t',\alpha'\right) & = &
-2N_r\rho\nonumber\\
 & & +t'\hat{E}_1\left(\rho_1=2\rho,\alpha_1=\alpha'/2\right)\nonumber\\
 & & +O(t'^2)
\label{eq3.13}
\end{eqnarray}
Here $\hat{E}_1(\rho_1,\alpha_1)$ is the ground state energy of the
``effective'' $t-J$ Hamiltonian (\ref{hat_h1}) on a chain with $N_r$ sites.
According to Eq.(\ref{eq2.14}) we can calculate the chemical potential
from the first derivative with respect to the charge density:
\begin{equation}
\mu_1(\rho,\alpha;t',\alpha')=-1+t' \hat{\mu}_1(\rho_1,\alpha_1)
\label{eq3.14}
\end{equation}
where
\begin{equation}
\hat{\mu}_1(\rho_1,\alpha_1)=\frac{1}{N_r}\frac{d \hat{E}_1}{d \rho_1}
\label{eq3.15}
\end{equation}
is the chemical potential of the $t-J$ model on a chain with $N_r$ sites and
$\alpha_1=\alpha'/2$ at $\rho_1=2\rho$.
\vspace{5mm}

(2) For $\rho\geq 1/2$ only rung states with $Q(x)=1(\sigma=\pm1)$
and $Q(x)=2$ are involved. From Eqs.
(\ref{eq3.7})-(\ref{eq3.10}) we see that the first part of the
effective Hamiltonian:
\begin{eqnarray}
\hat{H}_2(\alpha') & = &  \sum_{x=1}^{N_r-1}\hat h_2(x,x+1,\alpha')
\label{hat_h2}
\end{eqnarray}
with couplings:
\begin{eqnarray}
\hat h_2(x,x+1,\alpha') & = & \{-P_{21}(x,x+1)+\frac{\alpha'}{2}Q_{11}(x,x+1)
\nonumber\\
 & & +D_2(x,x+1)\}\label{eq3.16}
\end{eqnarray}
is indeed a $t-J$ model, if we treat electron
pair states $|Q(x)=2\rangle$ as {\it quasi-holes}.
The third term ($D_2$) on the right-hand side of (\ref{eq3.16}):
\begin{eqnarray}
\langle Q(x),Q(x+1)|\hat h_2(x,x+1)|Q(x),Q(x+1)\rangle \nonumber \\
=\Biggl\{
    \begin{array}{c}
       \hspace{2mm}0 \hspace{12mm} Q(x)=1, \;\;\;Q(x+1)=1\\
       -\frac{\alpha'}{2} \hspace{10mm}  Q(x)=2,\;\;\; Q(x+1)=1\\
       -\alpha' \hspace{10mm}  Q(x)=2,\;\;\; Q(x+1)=2\\
    \end{array}\nonumber\\
 & & \label{eq3.17}
\end{eqnarray}
takes into account the non-vanishing diagonal terms (\ref{eq3.10}).
The ground state energy $\hat{E}_2(\rho_2,\alpha_2)$ of the effective
Hamiltonian
\begin{equation}
\hat H_2=\sum_{x} \hat h_2(x,x+1,\alpha')
\label{eq3.18}
\end{equation}
on a chain of $N_r$ sites fixes the first order perturbation correction
to the ground state energy of the two leg ladder system:
\begin{eqnarray}
\label{eq3.19}
E\left(\rho, t=1, \alpha; t',\alpha'\right) & = &
 -2N_r[1-\alpha/2-\rho(1-\alpha)]\nonumber\\
 & & +\frac{t'}{2}\hat E_2(\rho_2,\alpha_2)+O(t'^2)\nonumber\\
 & &
\end{eqnarray}
where
\begin{equation}
\rho_2=2(1-\rho) \;\;,\;\; \alpha_2=\alpha'\,.
\label{eq3.20}
\end{equation}
Finally, we get the following relation between the chemical potential
$\mu(\rho,\alpha;t',\alpha')$ of the two leg ladder
\begin{equation}
\hat \mu_2(\rho_2,\alpha_2)=\frac{1}{N_r}\frac{d \hat E_2}{d \rho_2}\,,
\label{eq3.21}
\end{equation}
and for the effective Hamiltonian (
\ref{eq3.18}) on a chain
of $N_r$ sites:
\begin{equation}
\mu_2(\rho,\alpha;t',\alpha')=1-\alpha-
\frac{t'}{2}\hat \mu_2(\rho_2,\alpha_2)\,.
\label{eq3.22}
\end{equation}

Combining (\ref{eq3.14}) and (\ref{eq3.22}), we get for the width of the charge
density plateau at $\rho=\frac{1}{2}$ in the first order perturbation theory:
\begin{eqnarray}
W(\alpha,t',\alpha') & \equiv & \mu_2-\mu_1\,\hspace{3.0cm}\nonumber\\
 & = &\,2-\alpha  
-t'\left[\hat \mu_2(\rho_2=1,\alpha')/2\right.\nonumber\\
 & & \left.+\hat \mu_1(\rho_1=1,\alpha'/2)\right]
\label{eq3.23}
\end{eqnarray}


\section{Charge density plateaux on a three leg ladder\label{sec3}}

\begin{figure}[ht!]
\centerline{\includegraphics[width=8.0cm,angle=0]{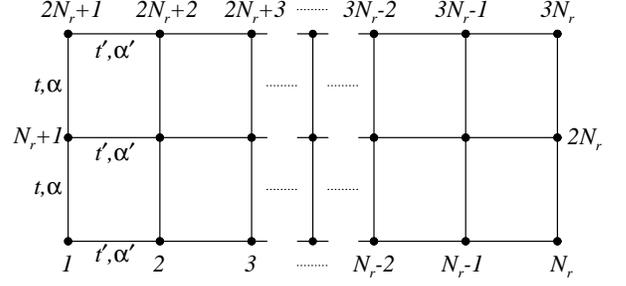}}
\caption{Structure of a three leg ladder with open boundary conditions
and $3N_r$ sites.}
\label{fig2}
\end{figure}

The perturbation treatment of the leg couplings will be applied now
on the three leg ladders. The geometry and the notion of states can be
seen in Fig.(\ref{fig2}).

The Hamiltonian (\ref{eq1.1}) for the 3-leg ($n=3$) case is again 
constructed from $t-J$ couplings
(\ref{eq2.4}) on the rungs and the legs, respectively, and is
given by the components
\begin{eqnarray}
h_r^{[3]}(x,\alpha) & = & h(x,x+N_r,\alpha)+h(x+N_r,x+2N_r,\alpha)
\,,\nonumber\\
h_l^{[3]}(x,\alpha') & = & \sum_{l=1}^3 h(x+(l-1)N_r,x+1+(l-1)N_r,\alpha')
\nonumber\\
 & & \label{eq4.1}
\end{eqnarray}

\subsection{0th order perturbation theory in $t'$}

The lowest energy eigenstates on the rung Hamiltonian $h_r^{[3]}(x,\alpha)$
have been calculated in [\onlinecite{kagan99}].
\begin{eqnarray}
|Q(x)=0\rangle & = & |0,0,0\rangle\nonumber\\
|Q(x)=1\rangle & = & \frac{1}{2}\{|\sigma,0,0\rangle+\sqrt{2}|0,
\sigma,0\rangle
+|0,0,\sigma\rangle\}\nonumber\\
|Q(x)=2\rangle & = & \frac{1}{\sqrt{4+2b^2}}\{|0,+,-\rangle -|0,-,+
\rangle \nonumber \\
 & & +b |+,0,-\rangle-b|-,0,+\rangle\nonumber\\
 & & +|+,-,0\rangle -|-,+,0\rangle\}\nonumber\\
|Q(x)=3\rangle & = &\frac{1}{\sqrt{6}}\{|\sigma,\sigma,-\sigma\rangle 
-2|\sigma,-\sigma,\sigma\rangle 
\nonumber \\
& & +|-\!\sigma,\sigma,\sigma\rangle\}\nonumber\\
 & & \label{eq4.4_9}
\end{eqnarray}
with
\begin{eqnarray}
\label{eq4_b}
b & = & \frac{-2}{\varepsilon_2}=(\sqrt{\alpha^2+8}-\alpha)/2\,,\\
\sigma & = & \pm 1\,.\nonumber
\end{eqnarray}
We have four types of states with charges $Q(x)=0,1,2,3$ respectively.
Even charge states with $Q(x)=0,2$ carry total spin $0$ 
and look like ``composite bosons''. Odd charge states with $Q(x)=1,3$
have total spin 1/2 and look like ``composite fermions''.
The energies of the four states are:
\begin{eqnarray}
\varepsilon_0 & = & 0,\quad\varepsilon_1=-\sqrt{2} \nonumber \\
\varepsilon_2 & = & -\frac{1}{2}(\sqrt{\alpha^2+8}+\alpha),\quad
\varepsilon_3=-\frac{3}{2}\alpha\,.
\label{eq4.10}
\end{eqnarray}
If we compute the ground state energy of the three leg ladder in the limit
of vanishing leg coupling $t'=0$, we have to discriminate the following three
cases:
\vspace{5mm}

(1) regime:\, $0\leq\rho\leq 1/3$
\quad($\rho=1/3-2q/3N_r$)

The ground state is a direct product of $N_r-2q$ rung states
$|Q(x)=1(\sigma)\rangle$ with charge $1$ and $2q$ rung states $|Q(x)=0\rangle$
with charge $0$.
\begin{equation}
E_1^{(0)}(\rho,\alpha)=(N_r-2q)\varepsilon_1+2q\varepsilon_0=3N_r\rho\varepsilon_1
\label{eq4.12}
\end{equation}
The creation of a charge $2$ state ($|Q(x)=2\rangle$)
and a charge $0$ state ($|Q(x)=0\rangle$) from two charge 1 states
costs energy, since
\begin{equation}
\varepsilon_0+\varepsilon_2 > 2\varepsilon_1 \hspace{5mm} \mbox{for} \hspace{5mm}\alpha < \frac{3}{\sqrt{2}}
\label{eq4.13}
\end{equation}

(2) regime:\, $1/3\leq\rho\leq 2/3$
\quad($\rho=1/3+2q/3N_r$)

The ground state is a direct product of $N_r-2q$ rung states with charge
$1$ and $2q$ rung states with charge $2$.
\begin{eqnarray}
E_2^{(0)}(\rho,\alpha)&=&(N_r-2q)\varepsilon_1+2q\varepsilon_2\nonumber \\
&=&N_r\{(2-3\rho)\varepsilon_1+(3\rho-1)\varepsilon_2\}
\label{eq4.14}
\end{eqnarray}
The creation of a charge $3$ state
and a charge $1$ state from two charge 2 states costs energy since
\begin{equation}
\varepsilon_1+\varepsilon_3 \geq 2\varepsilon_2 \hspace{5mm}  \mbox{for} \hspace{5mm}\alpha >0
\label{eq4.15}
\end{equation}
\vspace{5mm}

(3) regime:\, $2/3\leq\rho\leq 1$
\quad($\rho=1-2q/3N_r$)

The ground state is a product of $N_r-2q$ rung states with charge 3
and $2q$ with charge 2.
\begin{eqnarray}
E_3^{(0)}(\rho,\alpha)&=&(N_r-2q)\varepsilon_3+2q\varepsilon_2(\alpha) \nonumber \\
&=&N_r\{(3\rho-2)\varepsilon_3+3(1-\rho)\varepsilon_2\}
\label{eq4.16}
\end{eqnarray}
Note that  the ground state energies are linear again in the charge density.
Therefore, we get for the chemical potentials.
\begin{eqnarray}
\mu(\rho) & = & \frac{1}{3N_r}\frac{d E}{d \rho}
\label{eq4.17}\\
\mu_1^{(0)}(\rho,\alpha) & = & \varepsilon_1 \hspace{12mm} \mbox{for} \hspace{6mm}
0\leq\rho\leq 1/3\,,\nonumber\\
\mu_2^{(0)}(\rho,\alpha) & = & 
\varepsilon_2-\varepsilon_1 \hspace{5mm} 
\mbox{for} \hspace{5mm}
1/3\leq\rho\leq 2/3\,,\hspace{0.5cm}
\label{eq4.19}\\
\mu_3^{(0)}(\rho,\alpha) & = & \varepsilon_3-\varepsilon_2 \hspace{5mm} 
\mbox{for} \hspace{5mm}
2/3\leq\rho\leq 1\,.\nonumber
\end{eqnarray}
These are the results for the zeroth order ($t'=0$) in the
leg couplings [$h_l$ in (\ref{eq4.1})].

\subsection{1st order perturbation theory in $t'$}

The first order calculation starts from the matrix elements of the
leg couplings (\ref{eq4.1})
between the zeroth order states
(\ref{eq4.4_9}).
The results of this tedious calculation are summarized in
Table \ref{table1}.

\begin{table}[ht!]
\begin{tabular}{|c|c|c|c|}
  & & & \\[-13pt] \hline
$\rho$   &  $0\leq\rho\leq\frac{1}{3}$   & $\frac{1}{3}\leq\rho\leq\frac{2}{3}$
& $\frac{2}{3}\leq\rho\leq1$ \\
 & & & \\[-9pt] \hline
 & & & \\[-9pt]
$Q_{\mbox{rung}}$  & $Q=0, 1$ & $Q=1,2$  & $Q=2, 3$\\
 & & & \\[-9pt] \hline
 & & & \\[-9pt]
$\mu_{0th}$  & $\varepsilon_1$ & $\varepsilon_2-\varepsilon_1$  &
$\varepsilon_3-\varepsilon_2$ \\
 & & & \\[-9pt] \hline
 & & & \\[-9pt]
$\rho_{\mbox{eff.}}$  & $\rho_1=3\rho$   & $\rho_2=2-3\rho$
& $\rho_3=3\rho-2$  \\
 & & & \\[-9pt] \hline
 & & & \\[-9pt]
$H_{\mbox{eff.}}$  & $t_1 \hat H_1(\alpha_1)$  & $t_2 \hat H_2(\alpha_2)$
& $t_3 \hat H_3(\alpha_3)$  \\
 & & & \\[-9pt] \hline
 & & & \\[-9pt]
$t_{\mbox{eff.}}$   & $t_1=t'$
& $t_2=\frac{b^2+2\sqrt{2}b+2}{4(b^2+2)}t'$
& $t_3=\frac{3}{2(b^2+2)}t'$  \\
 & & & \\[-9pt] \hline
 & & & \\[-9pt]
$\alpha_{\mbox{eff.}}$  & $\alpha_1=\frac{3}{8}\alpha'$
& $\alpha_2=\frac{3(b^2+2)}{2(b^2+2\sqrt{2}b+2)}\alpha'$
& $\alpha_3=\frac{2(b^2+2)}{3}\alpha'$   \\
 & & & \\[-9pt] \hline
 & & & \\[-9pt]
$\mu_{1st}$  & $t_1\hat \mu_1(\rho_1, \alpha_1)$
& $-t_2\hat \mu_2(\rho_2, \alpha_2)$
& $t_3\hat \mu_3(\rho_3, \alpha_3)$   \\ \hline
\end{tabular}
\caption{Effective couplings for a three leg ladder}
\label{table1}
\end{table}

The first row defines the 3 regimes of charge density $\rho$.
In each regime, the zeroth order ground state is built up from direct products
of rung cluster states with charge Q listed in the second row.
The chemical potentials (\ref{eq4.17})-(\ref{eq4.19}) in zeroth order
are listed in the third row.

First order perturbation theory in the leg couplings leads to the
effective Hamiltonians with nearest neighbour interactions on a chain
of $N_r$ sites in each sector of $\rho$ (5-th row of Table \ref{table1}).
These Hamiltonians contain two parts:
\begin{eqnarray}
t_j\hat H_j(\alpha_j) & = & t_j\sum_{x=1}^{N_r-1} 
\hat h_j(x,x+1,\alpha_j)\,,\hspace{3mm}(j=1,2,3)\,,\nonumber\\
 & = & t_j[H_{t-J}(t=1,\alpha=\alpha_j)+D_j]\hspace{1mm}.
\label{eq4.21}
\end{eqnarray}
The first one is of the $t-J$ type (\ref{eq2.4}) with effective
hopping parameter $t_j$ (6-th row of Table \ref{table1}) and spin coupling
$\alpha_j$ (7-th row of Table \ref{table1}).

The second part $D_j, j=1,2,3$ with $D_1\equiv0$ takes into account
diagonal terms, which are not present in the $t-J$ Hamiltonian:

\begin{eqnarray}
\langle 1,2|\hat h_2(x,x+1,\alpha')
|1,2\rangle & = &
\frac{-(b^2+3)}{8(b^2+2)}\alpha'
\label{eq4.22}\\
\langle 2,2|\hat h_2(x,x+1,\alpha')
|2,2\rangle & = &
\frac{-(b^4+2b^2+3)}{2(b^2+2)^2}\alpha'\quad\quad
\label{eq4.23}\\
\langle 2,3|\hat h_3(x,x+1,\alpha')
|2,3\rangle & = &
\frac{-\alpha'}{2}
\label{eq4.24}\\
\langle 3,3|\hat h_3(x,x+1,\alpha')
|3,3\rangle & = &
\frac{-\alpha'}{2}
\label{eq4.25}
\end{eqnarray}

Note that in the three leg ladder case the effective hopping terms
$t_2$ and $t_3$ as well as the effective couplings $\alpha_2$ and
$\alpha_3$ depend on the rung coupling $\alpha$ via (\ref{eq4_b}).
This does not occur in the two leg ladder case.

If we denote the ground state energies of the effective Hamiltonian
$H_j$ on a chain with $N_r$ sites by $E_j(\rho_j, \alpha_j)$ and the
corresponding chemical potential by
\begin{equation}
\hat \mu_j(\rho_j, \alpha_j)=\frac{1}{N_r}\frac{d \hat E_j}{d \rho_j}
\label{eq4.26}
\end{equation}
we can express the first order correction to the chemical potential
of the three leg ladder in terms of (\ref{eq4.26}) (last row
of Table \ref{table1}). The relation between the charge density $\rho$ on the
ladder system and the charge density $\rho_j$ in the effective one
dimensional system can be found in the 4-th row of Table \ref{table1}.



\section{Numerical results\label{sec4}}

In this section we are going to present numerical results for the ground
state energies and the chemical potentials of the two and three
leg ladder.
Our results were obtained with open boundary conditions to
facilitate the comparison with future DMRG calculations which can
be done on larger systems. Other boundary
conditions -- e.g. periodic ones -- can be incorporated as well.


\subsection{Two leg ladders}

The ground state energies $\hat E_j(\rho_j,\alpha_j)$ of the effective
Hamiltonians $\hat H_j$ $j=1,2$ [cf. (\ref{hat_h1}) and (\ref{hat_h2})] on a
chain with $N_r$ sites have been computed for
$N_r=8,10,12,14,16,(18)$\footnote{The 18-site systems have been
evaluated for $Q=0,2,4,6,16,18$. For the $Q$ values inbetween, the dimension
of the Hilbert spaces exceeded our computer capacities.} and
\begin{eqnarray}
\alpha^{\prime}=2.7, & & \alpha_1=\frac{\alpha^{\prime}}{2}=1.35,\quad\alpha_2=
\alpha^{\prime}=2.7\,.
\end{eqnarray}
For $\rho_1=0$ and $\rho_2=0$ these energies are known:
\begin{eqnarray}
\label{e1_rho_0}
\hat E_1(\rho_1=0,\alpha_1) & = & 0\,,\\
\label{e2_rho_0}
\hat E_2(\rho_2=0,\alpha_2) & = & -\alpha_2(N_r-1)\,.
\end{eqnarray}
For $\rho_1=1$ and $\rho_2=1$ the ground state energies are given by
the nearest neighbour Heisenberg chain (with open boundary conditions)
\begin{eqnarray}
\label{e2_rho_1}
\hat E_1(\rho_1=1,\alpha_1=\alpha^{\prime}/2) & = & \frac{1}{2}
\hat E_2(\rho_2=1,\alpha_2=\alpha^{\prime})\,.
\end{eqnarray}
The finite-size dependence has been analyzed with an ansatz:
\begin{eqnarray}
\label{e0n_scal}
\hat e_j(\rho_j,\alpha_j) & = & \frac{\hat E_j(\rho_j,\alpha_j)}{N_r+N_j(\rho_j)}\,,
\quad\quad j=1,2\,,
\end{eqnarray}
for the ground state energies per site.

From (\ref{e2_rho_0}) we get
\begin{eqnarray}
\label{bound2_1}
N_2(\rho_j=0) & = & -1\,.
\end{eqnarray}
A finite-size analysis of the Heisenberg chain ($t-J$ chain at $\rho=1$)
yields
\begin{eqnarray}
\label{bound2_2}
N_j(\rho_j=1) & = & -0.6\,,\quad\quad j=1,2\,,
\end{eqnarray}
which means that finite-size effects change with $\rho_j$. We
assume here a linear interpolation:
\begin{eqnarray}
\label{bound2_3}
N_j(\rho_j) & = & -1+0.4\rho_j\,,\quad\quad j=1,2\,,
\end{eqnarray}
between the boundary values (\ref{bound2_1}) and (\ref{bound2_2}).

\begin{figure}[ht!]
\centerline{\includegraphics[width=8.0cm,angle=0]{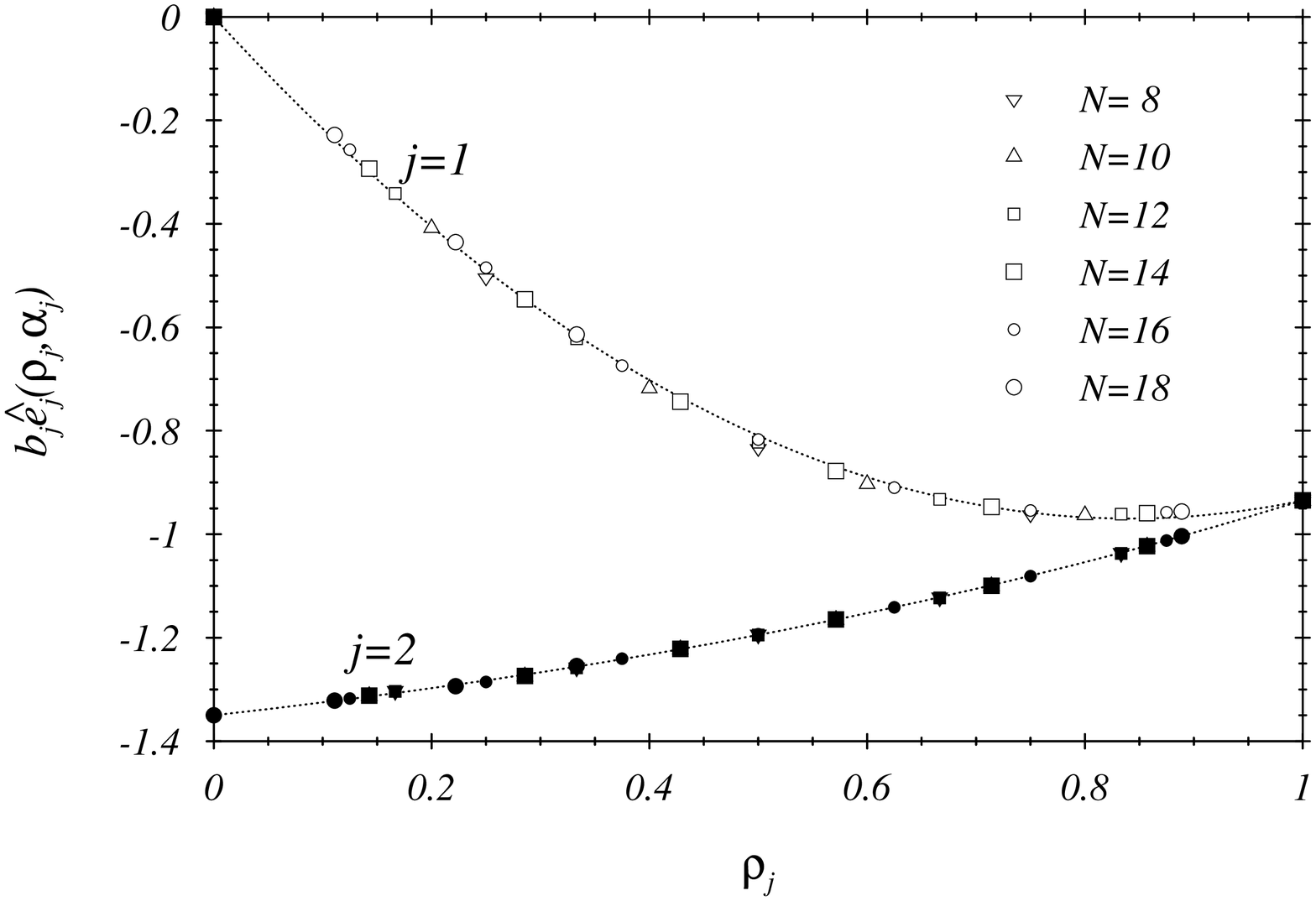}}
\caption{Ground state energies per site $b_j\hat e_j(\rho_j,\alpha_j)$
[Eqs.
(\ref{e0n_scal}) and (\ref{param_2})] for the effective Hamiltonians $\hat H_j$ 
($j=1,2$) of the
two leg ladder with $\alpha_2=\alpha'=2.7$ -- results for $N_r=8,10,
\ldots,(18)$ and optimized polynomial fits.}
\label{fig3}
\end{figure}

Indeed, this procedure has the effect, that the data points for
$N=12,14,16,(18)$ follow unique curves $\hat e_1(\rho_1,\alpha_1)$
and $\hat e_2(\rho_2,\alpha_2)/2$ as is demonstrated in Fig.\,\ref{fig3}.

The smooth dependence on the charge densities can be parametrized
in such a way
\begin{eqnarray}
\label{param_2}
b_j\hat e_j(\rho_j,\alpha_j) & = & b_j\hat e_j(\rho_j=0,\alpha_j) +\rho_j\left(a_j^{(0)}+
\right.\nonumber\\
 & & \left.a_j^{(1)}(1-\rho_j)+a_j^{(2)}(1-\rho_j)^2\right)
\end{eqnarray}
that the constraints (\ref{e1_rho_0}-\ref{e2_rho_1}) are built in
explicitely:
\begin{eqnarray}
\label{param_2b}
b_1=1.0\,, &  & \hat e_1(\rho_1=0,\alpha_1)=0,\nonumber\\
b_2=1/2\,, &  & \hat e_2(\rho_2=0,\alpha_2)=-\alpha^{\prime}\,.
\end{eqnarray}
In Table \ref{table_fit3} we list the coefficients of the
fit (\ref{param_2}).
\begin{table}[ht!]
\begin{tabular}{|c|c|c|c|}
  & & & \\[-13pt] \hline
 j & $a_j^{(0)}$   &  $a_j^{(1)}$   &  $a_j^{(2)}$ \\ \hline
\quad 1\quad \quad & \quad -0.9346\quad\quad  & \quad -1.3769\quad\quad &
\quad 0.0167\quad\quad \\ \hline
 2 &  \,0.4154  &  -0.2384 &  0.0580 \\ \hline
\end{tabular}
\caption{Parameters $a_j^{(0,1,2)}$, $j=1,2$ in the fit (\ref{param_2})
for the energies $b_j\hat e_j(\rho_j,\alpha_j)$ shown in Fig. \ref{fig3}.}
\label{table_fit3}
\end{table}

\begin{figure}[ht!]
\centerline{\includegraphics[width=8.0cm,angle=0]{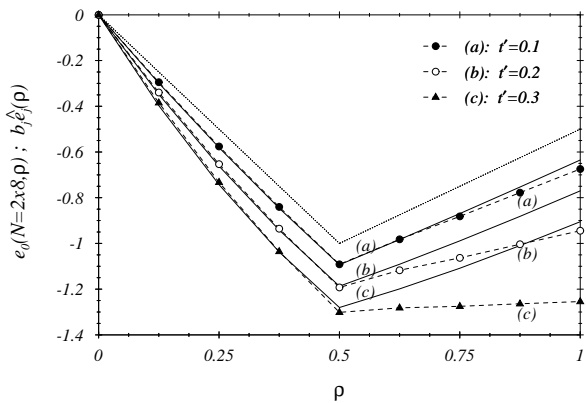}}
\caption{Zero (dotted line) and first order (solid lines) predictions
for two leg ladders with
$\alpha=0.5,\alpha'=2.7$ and leg couplings $t'=0.1(a),0.2(b),0.3(c)$
as well as the corresponding Lanczos energies for a $2\times 8$-site 
ladder (dashed lines).}
\label{fig4}
\end{figure}

In Fig. \ref{fig4} we compare the first order predictions
(\ref{eq3.13}), (\ref{eq3.19}) for the ground state energies per rung 
with a Lanczos diagonalization on a two leg ladder with 8 rungs
and couplings $\alpha=0.5$, $\alpha'=2.7$,
$t'=0.1(a),0.2(b),0.3(c)$. The charge density plateau at $\rho=1/2$
is clearly visible in the discontinuous change of the slope in the
energy per rung as function of $\rho$.

A remarkable agreement between Lanczos diagonalization (dashed curve)
and the perturbative result including first order corrections (solid
curve) is achieved for $\rho<1/2$. 
A comparison with the zeroth order result (dotted curve) demonstrates
that first order corrections are significant. 

The situation for
$\rho>1/2$ is different. Here, we observe deviations between the
Lanczos diagonalization (dashed curve) and the perturbative results
(solid curve) which increase monotonically with $t'$ and $\rho$.
Note, that the spin coupling $J'$ along the legs ($J'=t'\alpha'$)
is already close to $J'=1$ for $t'=0.3$ and $\alpha'=2.7$. This means
in particular for $\rho=1$, where the $t-J$ model reduces to a
Heisenberg model with spin couplings $J=0.5$ and $J'=0.81$
(for $t'=0.3$ and $\alpha'=2.7$), that the
product ansatz (\ref{eq3.1}) with rung cluster states (\ref{eq2.8})
is an inadequate starting point for a perturbative expansion.
We expect improvements, if we start with a direct product of more
complex clusters -- e.g. plaquettes on the two leg ladder.
Moreover, bound states (hole-pairs) may change the properties of
ground state at large charge density ($\rho \rightarrow 1$). This is
indeed the case for $\alpha>2$. \cite{troyer96}

\begin{figure}[ht!]
\centerline{\includegraphics[width=8.0cm,angle=0]{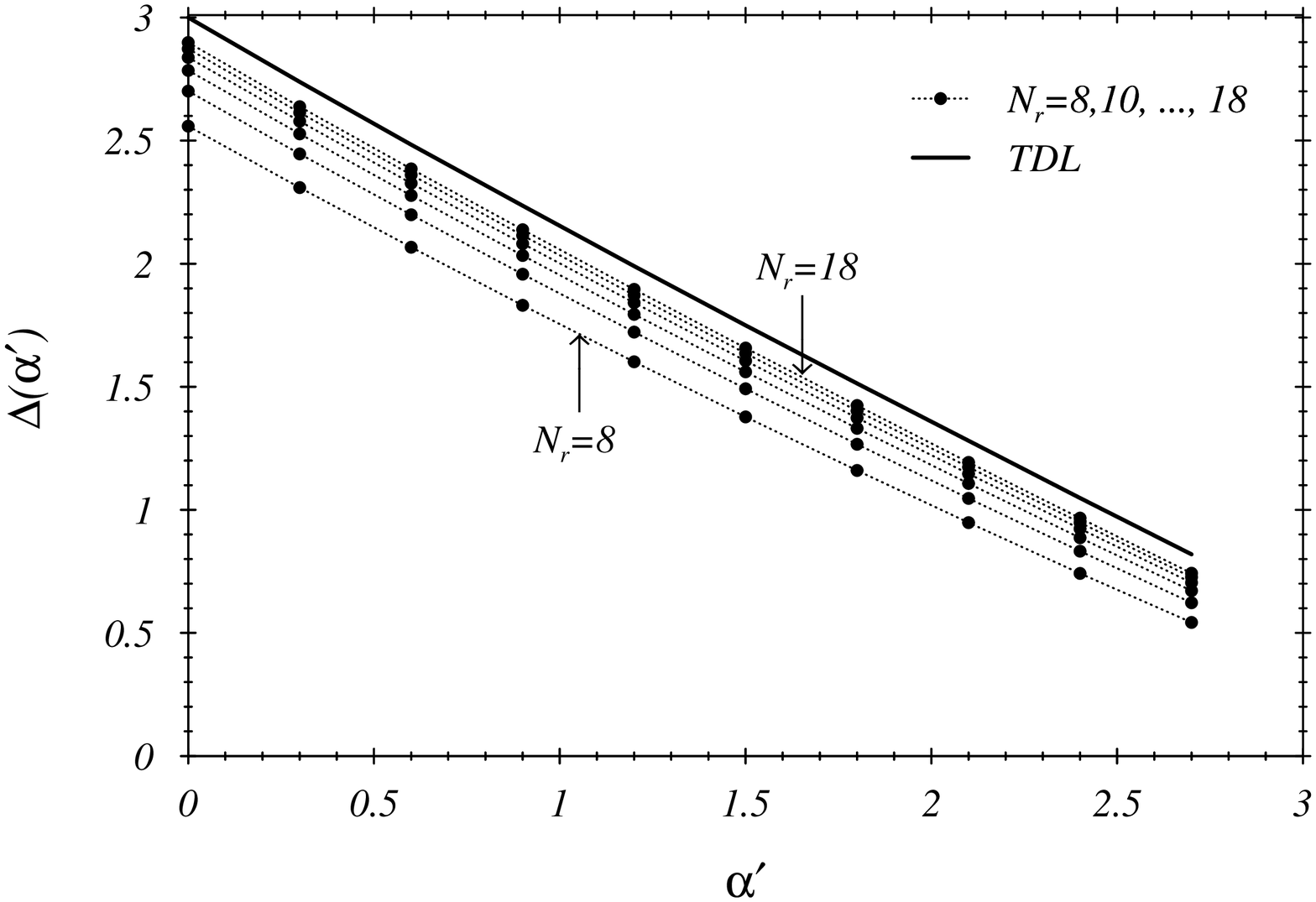}}
\caption{$\alpha'$-dependence (first order result) of the plateau width of
the chemical potential of a two leg ladder at $\rho=1/2$ -- finite
system results and BST-evaluation of the thermodynamical limit (TDL).}
\label{fig5}
\end{figure}

Let us next turn to the phase diagram of the two leg ladder,
which is defined by a vanishing plateau width $W(\alpha,t^{\prime},
\alpha^{\prime})$ [(\ref{eq3.23})]. In general one has to discuss
the phase boundary in the three dimensional parameter space 
$(\alpha,t^{\prime},\alpha^{\prime})$. In first order perturbation
theory in $t^{\prime}$ [cf. eq. (\ref{eq3.23})], however, it is
sufficient to discuss the boundary in the plane $(2-\alpha)/
t^{\prime}$ versus $\alpha^{\prime}$:
\begin{eqnarray}
\hspace{-0.5cm}
\frac{2-\alpha}{t^{\prime}} & = & \Delta(\alpha^{\prime})\nonumber\\ & = &
\frac{1}{2}\hat \mu_2(\rho_2=1,\alpha^{\prime})+\hat \mu_1(\rho_1=1,
\alpha^{\prime}/2)\,.\label{Delta_2}
\end{eqnarray}
We have determined the chemical potentials $\mu_1(\rho_1=1,\alpha^{\prime}/2)$,
$\mu_2(\rho_2=1,\alpha^{\prime})$ from a numerical calculation of the
ground state energies per site (\ref{e0n_scal}) on systems with
$N_r=8,10,12,14,16,(18)$ and
\begin{eqnarray}
\alpha^{\prime} & = & 0.0,\,0.3,\ldots,\,2.7\,.\nonumber
\end{eqnarray}
The $\alpha^{\prime}$-dependence of $\Delta(\alpha^{\prime})$
[right-hand side of (\ref{Delta_2})] is shown in Fig. \ref{fig5}.
There is a monotonic finite-size dependence and the thermodynamical
limit (solid curve) is estimated with the Bulirsch-Stoer (BST)
algorithm. \cite{bulirsch64}
The gapped phase with a nonvanishing plateau at $\rho=1/2$ is
characterized by
\begin{eqnarray}
\frac{2-\alpha}{t^{\prime}} & > & \Delta(\alpha^{\prime})
\end{eqnarray}
i.e. the formation of a plateau at $\rho=1/2$ is favoured for
\begin{itemize}
\item
small rung couplings $\alpha=J/t<2$
\item
large leg couplings $\alpha^{\prime}=J^{\prime}/t^{\prime}$
\end{itemize}

\subsection{Three leg ladders}

The ground state energies $\hat E_j(\rho_j, \alpha_j)$ of the effective
Hamiltonians [5th row of Table \ref{table1}, Eqs.(\ref{eq4.21})-(\ref{eq4.25})]
on a chain with $N_r$ sites have been computed for
$N_r=8,10,\ldots,16,(18)$ and
\begin{eqnarray}
\alpha=0.5, & & \alpha^{\prime}=2.7\\
\hspace{-0.0cm}\alpha_1=1.0125, & &
\alpha_2=2.041,\quad \alpha_3=6.132\,.\nonumber
\end{eqnarray}
The finite-size dependence has been analyzed with an ansatz of the
type (\ref{e0n_scal}) for the ground state energies per site
$\hat e_j(\rho_j,\alpha_j)$.

\begin{figure}[ht!]
\centerline{\includegraphics[width=8.0cm,angle=0]{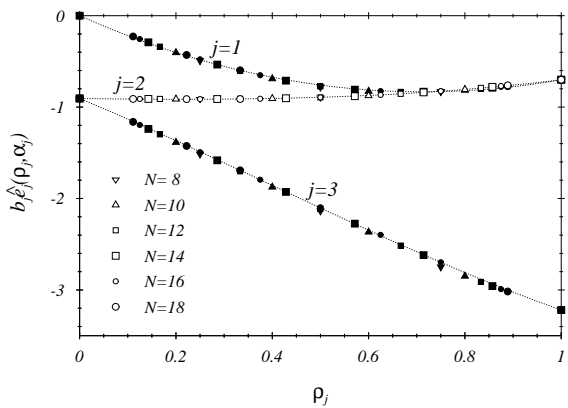}}
\caption{Ground state energies per site $b_j\hat e_j(\rho_j,\alpha_j)$
(\ref{e0n_scal},\ref{param_2}) for the effective Hamiltonians $\hat H_j$
($j=1,2,3$) of the
three leg ladder with $\alpha=0.5,\alpha'=2.7$
 -- results for $N_r=8,10,
\ldots,(18)$ and optimized polynomial fits.}
\label{fig6}
\end{figure}

The constraints at $\rho_j=0$
and at $\rho_j=1$
are taken into account in the linear interpolations
\begin{eqnarray}
N_j(\rho_j) & = & -1+0.4\rho_j\,,\quad\quad j=1,2\,,\nonumber\\
 & & \label{norm_3leg}\\
N_3(\rho_3) & = & -1+0.232\rho_3\,.\nonumber
\end{eqnarray}
The factors $b_j$ in Fig. \ref{fig6}, (\ref{param_3b})
\begin{eqnarray}
b_j & = & \frac{t_j}{t^{\prime}}\label{bj_3leg}
\end{eqnarray}
reflect the ``renormalization'' of the hopping term (6-th row in
Table \ref{table1}).
A plot of $b_je_j(\rho_j,\alpha_j)$, $j=1,2,3$, $\alpha=0.5$,
$\alpha^{\prime}=2.7$ on the finite systems 
is shown in Fig. \ref{fig6}.

The data points do not scatter with the system size but follow
unique curves, which can be considered as a reliable estimate
of the thermodynamical limit. Their dependence on the effective
charge densities $\rho_{\mbox{eff}}$  (fourth row in Table \ref{table1})
is parametrized by an ansatz of type (\ref{param_2}).
The resulting coefficients with
\begin{eqnarray}
\label{param_3b}
\hspace{-0.4cm}b_1=1.0\quad\quad & , & \hat e_1(\rho_1=0,\alpha_1)=0,\nonumber\\
b_2=0.49618 & , & \hat e_2(\rho_2=0,\alpha_2)=-\frac{8}{3}\alpha_2\frac{b^4+2b^2+3}
{2(b+2)^2}\,,\nonumber\\
b_3=0.44028  & , & \hat e_3(\rho_3=0,\alpha_3)=-\alpha_3\frac{b^4+2b^2+3}
{2(b+2)^2}\,,\nonumber\\
 & & 
\end{eqnarray}
are listed in Table \ref{table_fit6}.
\begin{table}[ht!]
\begin{tabular}{|c|c|c|c|}
  & & & \\[-13pt] \hline
 j & $a_j^{(0)}$   &  $a_j^{(1)}$   &  $a_j^{(2)}$ \\ \hline
\quad 1\quad \quad & \quad -0.7010\quad\quad  & \quad -1.6209\quad\quad & 
\quad -0.0260\quad\quad \\ \hline
 2 &  0.2054  &  -0.4364 &  0.1630 \\ \hline
 3 & -2.3129  &  -0.4460 &  0.5739 
 \\[-0pt] \hline
\end{tabular}
\caption{Parameters $a_j^{(0,1,2)}$, $j=1,2,3$ in the fits for the
energies $b_j\hat e_j(\rho_j,\alpha_j)$ shown in Fig. \ref{fig6}.}
\label{table_fit6}
\end{table}

Looking at the phase diagram of the three leg ladder we have
to distinguish four regimes in the three dimensional parameter
space of $\alpha,t^{\prime},\alpha^{\prime}$:
\begin{itemize}
\item[I]
\quad plateaux at $\rho=1/3,2/3$
\item[II]
\quad plateau at $\rho=1/3$, no plateau at $\rho=2/3$
\item[III]
\quad plateau at $\rho=2/3$, no plateau at $\rho=1/3$
\item[IV]
\quad no plateaux.
\end{itemize}
The phase boundaries are defined by the vanishing of the plateau width
(cf. rows 3 and 8 in Table \ref{table1}):
\begin{eqnarray}
\label{param_3}
\rho=1/3: & & \nonumber\\ 
W_{1/3} & = &\varepsilon_2-2\varepsilon_1-\\
 & &  t^{\prime}\left(b_2\hat \mu_2(\rho_2=1,\alpha_2)+
b_1\hat \mu_1(\rho_1=1,\alpha_1)\right)\nonumber\\
\rho=2/3: & & \nonumber\\ 
W_{2/3} & = & \varepsilon_3+\varepsilon_1-2\varepsilon_2+\\
 & & t^{\prime}\left(b_3\hat \mu_3(\rho_3=0,\alpha_3)+
b_2\hat \mu_2(\rho_2=0,\alpha_2)\right)\nonumber\,.
\end{eqnarray}
In first order perturbation theory the widths are linear in
$t^{\prime}$ and it is therefore convenient to represent the
phase boundaries in the form
\begin{eqnarray}
\label{phase_bound_3}
\hspace{-0.5cm}t^{\prime}_{1/3} & = & t^{\prime}_{1/3}(\alpha,
\alpha^{\prime})\nonumber\\
 & = & \frac{\varepsilon_2-2\varepsilon_1}{\left(b_1\hat \mu_1(\rho_1=1,\alpha_1)+
b_2\hat \mu_2(\rho_2=1,\alpha_2)\right)} \\[5pt]
\hspace{-0.5cm}t^{\prime}_{2/3} & = & t^{\prime}_{2/3}(\alpha,
\alpha^{\prime})\nonumber\\
 & = & -\frac{\varepsilon_3+\varepsilon_1-2\varepsilon_2}
{\left(b_3\hat \mu_3(\rho_3=0,\alpha_3)+
b_2\hat \mu_2(\rho_2=0,\alpha_2)\right)} \,.
\end{eqnarray}
The four phases defined above are characterized by
\begin{itemize}
\item[I]
\quad $t'_{1/3}(\alpha,\alpha^{\prime})
\geq t^{\prime}\geq 0$\,,\quad\quad
$t'_{2/3}(\alpha,\alpha^{\prime})\geq t^{\prime}\geq 0$\,,
\item[II]
\quad $t^{\prime}_{1/3}(\alpha,\alpha^{\prime})
\geq t^{\prime}\geq 0$\,,\quad\quad
$t^{\prime}\geq t'_{2/3}(\alpha,\alpha^{\prime})$\,,
\item[III]
\quad $t^{\prime}
\geq t^{\prime}_{1/3}(\alpha,\alpha^{\prime})$\,,\quad\quad\quad\quad
$t'_{2/3}(\alpha,\alpha^{\prime})\geq t^{\prime}\geq 0$\,,
\item[IV]
\quad $t^{\prime}
\geq t^{\prime}_{1/3}(\alpha,\alpha^{\prime})$\,,\quad\quad\quad\quad
$t^{\prime}\geq t'_{2/3}(\alpha,\alpha^{\prime})$\,.
\end{itemize}

\begin{figure}[ht!]
\centerline{\includegraphics[width=8.0cm,angle=0]{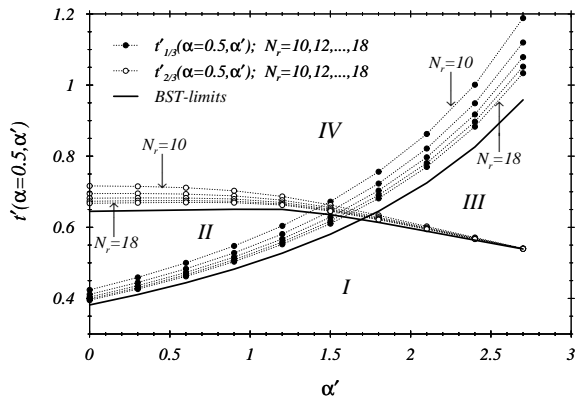}}
\caption{First order results for the $(t',\alpha')$-dependence
of the boundaries of the plateau regimes ($I-IV$) of a
three leg ladder -- finite system results and BST-evaluation 
of the TDL.}
\label{fig7}
\end{figure}

We have computed the phase boundaries $t^{\prime}_{1/3}(\alpha,
\alpha^{\prime})$ and $t^{\prime}_{2/3}(\alpha,\alpha^{\prime})$
from the ground state energies of the effective Hamiltonians
$\hat H_j$ (\ref{eq4.21}) on systems with $10,12,\ldots,(18)$ sites for 
\begin{eqnarray}
\alpha=1/2 & , & \alpha'=0.0,0.3,\ldots, 2.7\,.\nonumber
\end{eqnarray}
This projection of the phase diagram is shown in Fig. \ref{fig7}.
The thermodynamical limits (TDLs; solid curves) are estimated with
the BST algorithm. \cite{bulirsch64}

\section{Discussion and perspectives}

Quasi onedimensional quantum systems -- like the Heisenberg model
or the $t-J$ model -- defined on ladders with $l$ legs show a
characteristic sequence of gaps in their spectrum. They appear
as plateaux in the magnetization curve $M(B)$ and the charge
density $\rho(\mu)$ (as functions of the magnetic field $B$ and
chemical potential $\mu$, respectively) for the spin and charge
degrees of freedom.
The quantization rule of Oshikawa, Yamanaka and Affleck 
\cite{oshikawa97} defines
the values of $M$ and $\rho$ where these might occur.
The mechanism which creates the plateaux can be studied in a
perturbation expansion in the leg couplings $t'$; $\alpha'=J'/t'$ 
fixed. To zeroth order the ground states -- at fixed magnetization
$M$ and/or charge density $\rho$ -- are products of rung cluster
states. The latter can be classified by a cluster spin $S_3$ and
charge $Q$. First order perturbation theory leads to an effective
interaction between the rung cluster states.

We have studied in this paper the 0th and 1st order perturbation
expansion in the leg coupling $t'$ (for $\alpha'$ fixed) on two
and three leg ladders. The magnetization -- given by the total 
spin -- has been assumed to be zero. Our results on the two leg
ladder can be summarized as follows:

$i)$ The ground state in the two regimes
\begin{eqnarray}
\left.\begin{array}{cc}
\quad 0\leq\rho\leq 1/2 & \quad\quad (Q=0,1)\\   
1/2\leq\rho\leq 1\quad & \quad\quad (Q=1,2)
\end{array}\right.
 & & \nonumber
\end{eqnarray}
is built up from a direct product of rung cluster states with
$Q=0,1$ and $Q=1,2$, respectively.

Even charge states ($Q=0,2$) have total spin 0 and are bosonic,
odd charge states ($Q=1,(3$ for the 3-leg ladder$)$) have total
spin 1/2 and are fermionic.
For $\rho=0,1/2,1$ all rung cluster
states have the same charge $Q=0,1,2$, respectively.

$ii)$ The effective Hamiltonian, which describes the interaction
between the rung cluster states in first order perturbation
theory, looks like a generalized $t-J$ model on a chain, if we
treat fermionic states ($Q=1$) as electrons, bosonic ones
($Q=0,2$) as holes. A comparison of the perturbative results
with exact diagonalizations yields good agreement for the
ground state energies in the first regime ($\rho<1/2$, $t'\leq 0.3$),
but increasing deviations with $\rho$ and $t'$ in the second one.

$iii)$ Increasing the size of the cluster we started with, will
improve these results and might reveal further gaps in the spectrum.
In a next step we will consider a direct product of plaquette
cluster states on the two leg ladder (Fig. \ref{fig1}). Here,
cluster ground states with charges $Q(x)=0,1,2,3,4$ occur, which
might induce additional plateaux at $\rho=1/4$ and $\rho=3/4$.
Indeed, evidence for the existence of a long range charge density
wave state in the $t-J$ two leg ladder at $\rho=3/4$ has been found
in a DMRG calculation.\cite{white03}

$iv)$ We have studied the stability of the charge density plateau
and the phase diagram in the regime where first order perturbation
theory is applicable. We found that the formation of a charge 
density plateau is favoured for
\begin{itemize}
\item
small values of the couplings
$\alpha=J/t$ on the rungs
\item
large values of the couplings
$\alpha'=J'/t'$ on the legs.
\end{itemize}

In the three leg ladder case, the ground states in 0th order are
built up again from rung cluster states with two charges:
\begin{eqnarray}
\left.\begin{array}{cc}
\quad 0\leq\rho\leq 1/3 & \quad\quad (Q=0,1)\\  1/3\leq\rho\leq 2/3 & 
\quad\quad (Q=1,2)\\
2/3\leq\rho\leq 1\quad & \quad\quad (Q=2,3) 
\end{array}\right.
 & & \hspace{0.1cm}\nonumber
\end{eqnarray}
For $\rho=0,1/3,2/3,1$ all the rung clusters have the same charges:
$Q=0$, $Q=1$, $Q=2$, $Q=3$, respectively.

The method developped here for ladder systems should be applicable
whenever finite clusters containing the ``large'' couplings can be
defined in a natural way.
If we for instance consider the $t-J$ model on a Shastry-Sutherland lattice,
\cite{shastry81} the ``large'' couplings on the diagonals define
two site clusters. The same arguments we developped here for the
two leg ladder (with two site clusters) can be repeated and we
expect again a charge density plateau at $\rho=1/2$. Of course
the effective interaction between the cluster states depend on the
geometry of the lattice and are therefore different for the
Shastry-Sutherland lattice and the two leg ladder.

The method developped here for charge density plateaux is also applicable
to magnetization plateaux provided that finite clusters containing the
``large'' couplings can be defined in a natural way.
Again, the Heisenberg model on a Shastry-Sutherland lattice \cite{shastry81}
is a good example.

However, for a realistic description of the experimentally found
magnetization plateaux $M=S/N=1/6,1/8,1/16$ ($S$ total spin, $N$
total number of sites) one should start from clusters which at 
least -- requiring integer total spin for the clusters -- 
contain $6,8,16$ sites! Of course the computation of the
cluster ground states and in particular of their interaction
becomes more involved for decreasing magnetization values.


\acknowledgments

We are indebted to M. Karbach for a critical
reading of the manuscript.






\end{document}